\documentclass[a4paper,10pt]{article}

\usepackage[english]{babel}
\usepackage[latin1]{inputenc}
\usepackage{graphicx}
\usepackage{epsfig}
\usepackage{hyperref}
\usepackage{wrapfig}

\usepackage[normalem]{ulem}

\newcommand{\pr}{\partial}
\newcommand{\rta}{\rightarrow}

\newcommand{\ra}{\rangle}
\newcommand{\la}{\langle}

\title{On the failure of mean-field theories near a critical point}
\author{Navinder Singh\footnote{navinder.phy@gmail.com}\\ Physical Research Laboratory, Ahmedabad, India -380009}

\begin{document}

\maketitle

\begin{abstract}
It is well known that mean-field theories fail to reproduce the experimentally known critical exponents. The traditional argument which explain this failure of mean-field theories near a critical point is the Ginsburg criterion in which diverging fluctuations of the order parameter is the root cause. We argue, contrary to the  above mentioned traditional view, that diverging fluctuations in real physical systems near a critical point are genuine consequence of the breakdown of the property of statistical independence, and are faithfully reproduced by the mean-field theory. By looking at the problem from the point of view of "statistical independence" the divergence of fluctuations in real physical systems near criticality becomes immediately apparent as a connection can be established between diverging correlation length and diverging fluctuations. To address the question of why mean-field theories, much successful qualitatively, fail to reproduce the known values of critical indices we argue, using the essential ideas of the Wilsonian renormalization group, that mean-field theories fail to capture the long length scale averages of an order parameter near a critical point.       
\end{abstract}

\section{Divergence of fluctuations near a critical point in mean-field theories: breakdown of statistical independence}

The contradiction of Mean-Field Theories (MFTs) with experiment near a critical point was apparent, even from the days of Andrews (1869)\cite{1}. Problems of MFTs were well established by 1960s and were resolved by the advent of renormalization group theory due to Wilson and others in early 1970s\cite{2,3}. MFT fails to predict correct values of critical indices, for example,  the exponent $\beta$ of magnetization $M \propto (T-T_c)^\beta$ differs in MFT and in experiment:  $\beta_{theory}=\frac{1}{2}$ and $\beta_{experiment}=\frac{1}{3}$. One can generally say that MFTs are qualitatively correct but quantitatively wrong. In the traditional viewpoint,  this failure of MFT is assigned to diverging fluctuations near a critical point\cite{4}. One generally argues as follows. Consider Landau's formulation of MFTs\cite{5} and write free energy per unit volume in the presence of magnetic field $H$ as
\[g(T,M) = g_0+ a (T-T_c) M^2 + u(T) M^4- M H.\]
As magnetization per unit volume $M$ is a thermodynamical variable, probability of its fluctuation, from $M$ to
$M+\delta M$ with respect to unfluctuated state is:
\[\frac{w}{w_0} = \frac{e^{-\frac{g(T,M+\delta M)}{k_B T}}}{e^{-\frac{g(T,M)}{k_B T}}}.\]

By Taylor expanding free energy about unfluctuated value, one will have

\[\frac{w}{w_0} = e^{-\delta M \frac{1}{2 k_B T}\frac{\pr^2 g}{\pr M^2}}\]

As $\chi$ the magnetic susceptibility is defined as $\frac{\pr M}{\pr H}|_{T=T_c}$,
differentiating the free energy (above) twice wrt $M$ one can write $\chi$ as

\[\chi = 1/(\frac{\pr^2 g}{\pr M^2}|_{T\rta T_c})\propto  \frac{1}{T-T_c}.\]

This gives the root-mean-square (rms) value ($rms_M$) of $M$ at $T \rta T_c$:
\[rms_M  \propto \chi \propto \frac{1}{T-T_c}.\]
As the fluctuations are Gaussian fluctuations. Thus, one notices that as $\chi$ diverges at $T_c$, the rms fluctuation in $M$ also diverges! OR within mean-field theory near the critical point, fluctuations dominate on the average behavior. This diverging fluctuations near a critical point is viewed as a problem of MFTs, and thus MFTs predict  wrong values of the critical indices. OR in more appropriate language MFT predicts its own demise\cite{4}.

Contrary to this we argue that diverging fluctuations in real physical systems near a critical point are genuine (they are actually present in real physical systems\footnote{Critical opalescence is a concrete example\cite{6}.}) and the MFT theory faithfully reproduce that. The prediction of diverging fluctuations is an other success of MFTs (including its qualitative description of various phases). The reason  why MFTs fail to predict correct values of critical indices is different one and is explained below using Wilsonian renormalization group ideas.

To support our above argument of faithful reproduction of diverging fluctuations by MFT near a critical point we give two justifications. First one is based on the breakdown of a very important property of Statistical Independence (SI) when the correlation length diverges near a critical point. From the breakdown of SI  one can show the divergence of fluctuations. The other one is based on the fluctuation-dissipation theorem and power law correlation functions.

Divergence of correlation length can be seen very easily from Ginzburg-Landau formulation\cite{3} in which order parameter $M(x)$ vary weakly on atomic dimensions:
\[F = \int d^3 x \{ (\nabla M(x))^2 + r(T) M(x)^2 + u(T) M(x)^4 - B(x) M(x)\}.\]
One can calculate the correlation length $\xi$ (length scale over which the effect of a fixed test spin extends
out to other spins) by considering a delta function magnetic field: $B(x) = B_0 \delta^3(x)$.  By standard procedure of minimizing $F$ one  can easily show\cite{3}:
\[-\nabla^2 M(x) + r(T) M(x) = B_0 \delta^3(x).\]
Above differential equation can be solved easily with Fourier transforms: $ M(x) \sim \frac{e^{\sqrt{r(T)} |x|}}{|x|}.$
 And the correlation length can be defined: $\xi = \frac{1}{r(T)} \propto
\frac{1}{(T-T_c)^{1/2}}$. The correlation length diverges at the critical point, a standard result.

The diverging correlation length is intimately connected to diverging fluctuations. This can be very clearly seen with the breakdown of Statistical Independence (SI). The fundamental reason is that due to large correlation length various sub-parts of the system respond in a correlated way. This renders the system non-self-averaging! The fundamental hypothesis  of SI (actually an important property of ordinary statistical mechanical systems, i.e., systems with short range inter-particle interactions) as advocated by Landau\cite{5} breaks down and fluctuations in the sum-function observables ($f = \sum_i f_i$) do not obey $\frac{1}{\sqrt{N}}$ law\cite{5} near criticality. Or fluctuations do not converge:

Let a given system is divided into two parts. Let $\rho_1$ be the distribution function for part 1 and $\rho_2$ for part 2. If the two parts of the system are correlated (due to large correlation length), then

\[\rho_{12} \ne \rho_1 \rho_2.\]

Where $\rho_{12}$ is the distribution function of whole system. This is the breakdown of SI. One immediate consequence is that statistical average of two physical quantities will obey:

\[\la f_{12}\ra \ne \la f_1\ra \la f_2\ra.\]

If $f =\sum_{i=1}^N f_i$ (system containing $N$ sub-parts), then
\[\la (\Delta f)^2 \ra  =  \la (f-\la f\ra)^2\ra = \la\left(\sum_i \Delta f_i\right)^2\ra \ne \sum_i \la(\Delta f_i)^2\ra.\]

As $\la \Delta f_i \Delta f_j \ra \ne \la \Delta f_i\ra \la \Delta f_j\ra$. From this it follows that the relative fluctuation ($\frac{\sqrt{\la (\Delta f)^2 \ra}}{\la f \ra}$) does not obey $\frac{1}{\sqrt{N}}$ (whereas in standard statistical mechanical systems due to large $N$ fluctuations are negligible). Thus systems near criticality, due to the presence of long correlation length, lose the important property of self-averaging!
 
{\it From above we observe a very important connection: diverging correlation length and breakdown of SI are connected with each other.} Diverging fluctuations due to diverging correlation length can also be seen  through fluctuation-dissipation theorem:

By Fluctuation-Dissipation Theorem (FDT) the magnetic susceptibility $\chi$ and the correlation function $\Gamma(r)=\la M(r) M(0)\ra - \la M(r)\ra\la M(0)\ra$ are related to each other\cite{7}:

\[\chi =\frac{1}{k_B T} \int d^3 r \Gamma(r).\]

At the critical point the correlation function assumes a power law as the correlation length diverges ($\Gamma(r) \sim \frac{1}{r} e^{-r/\xi}$). Due to the power law correlation function the above integral do not converge and thus susceptibility diverges which from above analysis of Gaussian fluctuations leads to diverging rms value of $M$ ($rms_M$). 

Above two justifications one based upon SI and the other on FDT do not involve mean-field ideas and thus independently verify what is predicted by the MFTs and seen in actual experiments.

\section{Why does MFTs fail to reproduce the experimental values of critical indices?}

To understand why MFT fails to reproduce critical data one has to consider the construction of MFT.  Consider the  standard Ising model (figure ~\ref{ising}):
\begin{figure}[h!]
\centering
\begin{tabular}{cc}
\includegraphics[height = 4cm, width =6cm]{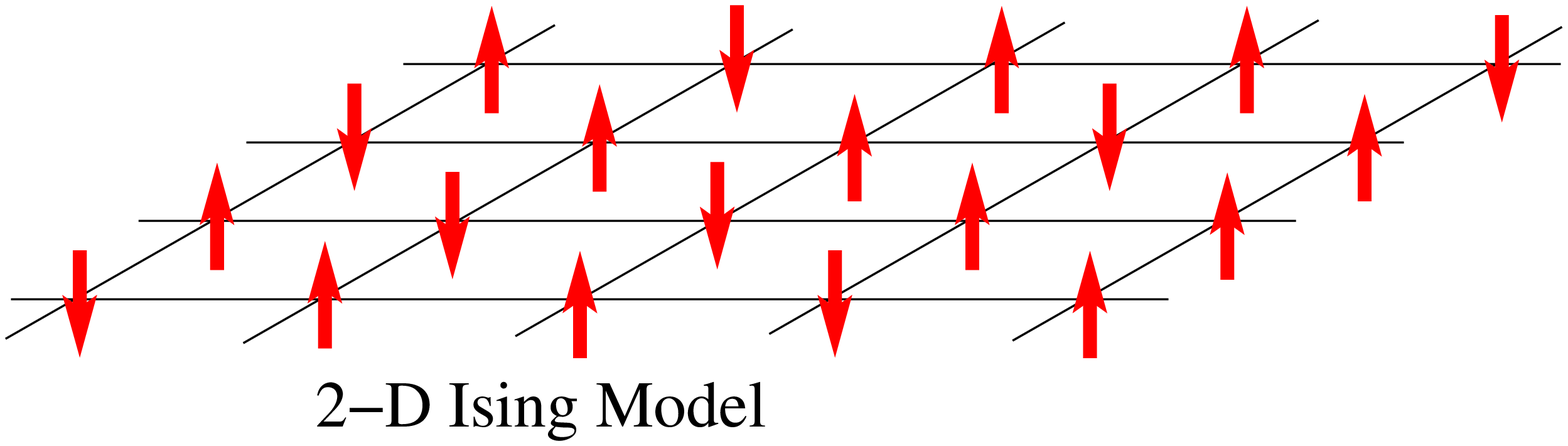}&
\includegraphics[height = 5cm, width =5.5cm]{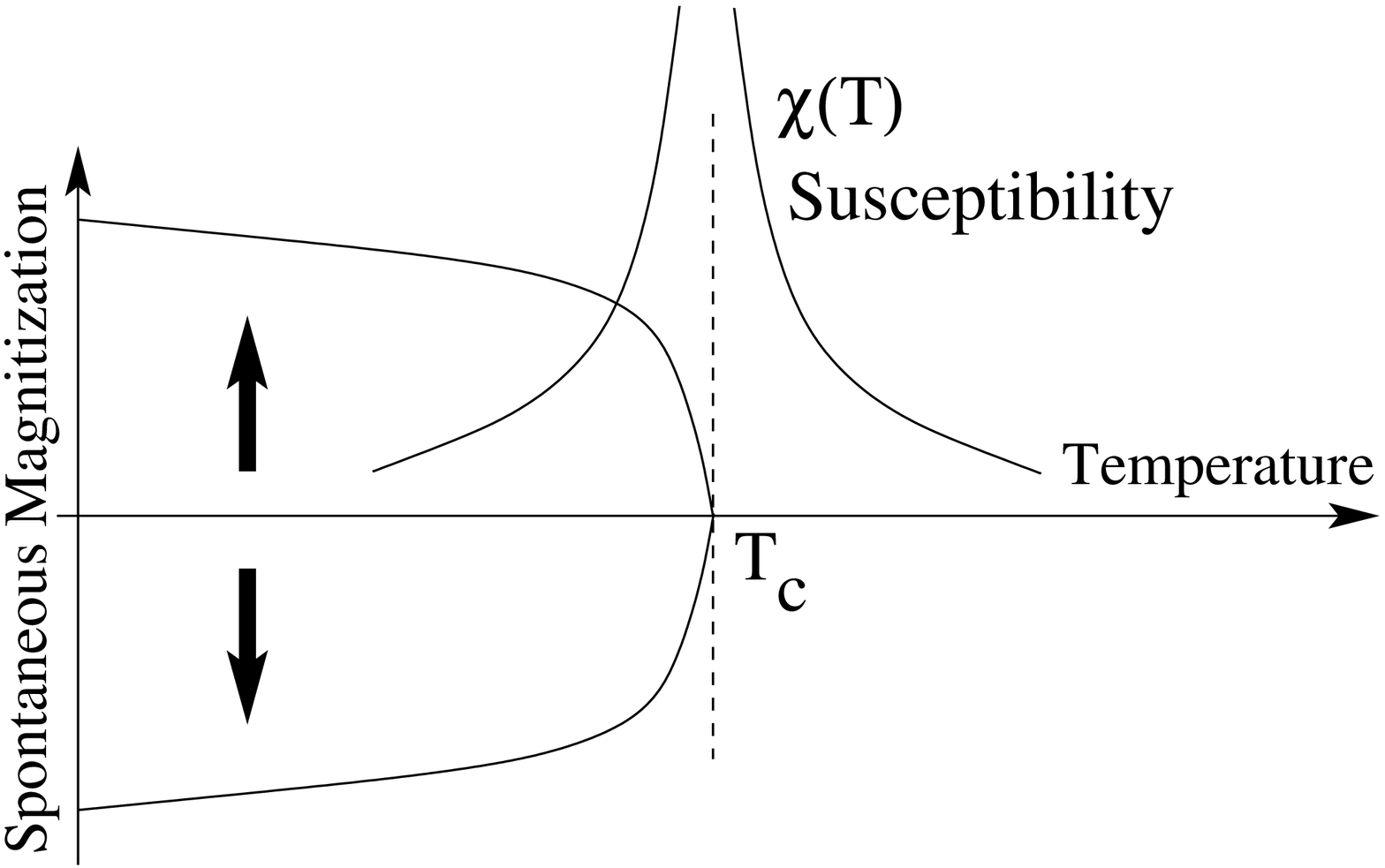}
\end{tabular}
\caption{(a) 2-D Ising model (b) Phase diagram of the model.}
\label{ising}
\end{figure}
Let the system be immersed in an external magnetic field $H$. The Hamiltonian in dimensionless form ($K =\frac{J}{k_B T}$ and $h= \frac{\mu H}{k_B T}$) is:
\[\frac{H}{k_B T} = - K \sum_{<nn>} \sigma_i \sigma_J - h \sum_i \sigma_i.\]
There are two competing tendencies (1) lining-up of spins due to $K$ and $h$, and (2) disruption due to thermal agitation. At low temperatures first prevails and at high the second. Ising model Mean-field theory is done in standard way\cite{1,4}. First, consider only one spin  immersed in magnetic field. Ensemble average is:
\[\la \sigma \ra = \frac{\sum_{\sigma = \pm 1} \sigma e^{h\sigma}}{\sum_{\sigma
= \pm 1} e^{h \sigma} } = \tanh(h).\]
Next, consider the immersed test spin interacting with many neighboring spins:
\[\la \sigma \ra = \frac{\sum_{\sigma = \pm 1} \sigma e^{h_{total} \sigma}}{\sum_{\sigma
= \pm 1} e^{ h_{total} \sigma}} = \tanh(\underbrace{h + K \sum_{<nn>}\sigma_i}_{h_{total}}) \simeq \tanh(h_{eff}).\]
With
\[h_{eff} = h + {\rm effective~field~due~to~other~spins}= h+ z  K ~\la \sigma \ra .\]

Here the exact field $K \sum_{<nn>} \sigma_i$ due to nearest neighbor spins is
replaced by an effective field $\sim z K \la \sigma \ra$ ($z$ is called the
coordination number (number of nearest spins). This constitutes the MF approximation.

Right here one makes an error when one is near the critical point: Near the critical point fluctuations
become long ranged (diverging correlation length). These long ranged correlations enhance the effective field seen by our test spin. Thus
\[ h_{eff} > h + k \sum_{<nn>} \sigma_i.\] Or the MFTs fail to account for the effective field generated due to weak non-zero values of magnetization on a much larger length scales (as compared to lattice constant) when temperature is only slightly less than the critical temperature. In other words in MFTs averages are done on a length scale (of the order of lattice constant) much smaller than the correlation length which is quite large as compared to lattice constant near criticality. And averages on a smaller length scale gives $\la\sigma\ra \simeq 0$ when temperature is slightly less than the critical temperature, but on a larger length scale of the order of correlation length it is not zero. To properly take into account this important   long distance effects one must average out fluctuations on all length scales step by step (see for example\cite{3}). One very visual procedure is the Kadanoff blocking method\cite{3} and results are in very good agreement with experiment and with exact solutions in special cases\cite{8}. See also the last section in\cite{6}.

\section{Conclusion}

A common misconception that mean-field theory predicts its own demise is clarified. Near a critical point diverging fluctuations are {\it actually} present in real physical systems, and the MFTs very faithfully reproduce those. Thus, this should be viewed as another success of MFTs, not their failure. The reason why MFTs predict wrong values of critical indices is that the MFTs fail to account for weak non-zero values of order parameter on a much larger length scales (as compared to lattice constant) when temperature is only slightly less than the critical temperature. In other words in MFTs averages are done on a length scale much smaller than the correlation length which is quite large as compared to lattice constant near criticality.

\end{document}